\documentstyle[preprint,aps]{revtex}
\begin{document}
\tightenlines
\title{Strong light fields coax intramolecular reactions on femtosecond time scales}
\author{M. Krishnamurthy, F. A. Rajgara, and D. Mathur}
\address{Tata Institute of Fundamental Research, 1 Homi Bhabha Road, 
Mumbai 400 005, India.}
\date{\today}
\maketitle
\begin{abstract}
Energetic H$_2^+$ ions are formed as a result of intra-molecular rearrangement during fragmentation of linear alcohols (methanol, ethanol, propanol, hexanol, and dodecanol) induced by intense optical fields produced by 100 fs long, infrared, laser pulses of peak intensity 8$\times$10$^{15}$ W cm$^{-2}$. Polarization dependent measurements show, counterintuitively, that rearrangement is induced by the strong optical field within a single laser pulse, and that it occurs before Coulomb explosion of the field-ionized multiply charged alcohols.  
\end{abstract}
\maketitle
	Molecular rearrangements are a very important facet of chemical dynamics and play a key role in understanding reaction pathways \cite{gen1}. Molecular rearrangements are basically  unimolecular reactions in which the energy absorbed by the molecule is channeled to certain vibrational modes that leads to reorientation or realignment of the molecular structure. The molecule with the altered morphology is then able to evolve along reaction pathways that were otherwise not available, and this results in a particular set of reaction products. Unimolecular reactions, of which isomeric transformation are a subset, play a vital role in many important reactions in nature, like photoabsorption in rhodopsin that is responsible for vision \cite{br}. Many rearrangements, like the McLafferty rearrangement \cite{lafferty}, often play a pivotal role in analytical chemistry in that they provide deeper insights into molecular fragmentation and reaction mechanisms. The advent of femtochemistry \cite{zewail} has enabled molecular motion to be deciphered in very direct fashion, and has provided new insights into intramolecular motion and reaction pathways that has enabled the subject to make qualitative progress on a number of fronts. 

	{\it A priori}, it might be considered perfectly legitimate to postulate that rearrangement reactions are of little or no concern in situations where molecules are irradiated by ultrashort laser light whose intensity is large enough so that the associated electric fields start to match intramolecular Coulombic fields. Light intensities of 10$^{16}$ W cm$^{-2}$ achieve this situation and make optical field-induced multiple ionization inevitable in most polyatomic molecules. The rapid deposition (on femtosecond timescales) of laser energy that gives rise to ejection of more than one molecular electron also leads to efficient, and fast fragmentation of molecules via a Coulomb explosion. Most often, the molecular ion peak that is measured in a mass spectrum in such instances is negligibly small compared to the yield of fragments ions \cite{int-gen1}. This is even more so in measurements that are made in ``intensity-selected" mode \cite{sudeep} wherein only the central (most intense) part of the laser focal volume is sampled and contribution to the mass spectrum from the lower intensity regions of the focal volume is avoided. In such intense field situations, conventional wisdom would dictate that it is very improbable that the multiply charged molecular ion that is formed will undergo re-orientation so as to lead to rearrangement products.  Multiple ionization in the Franck-Condon region leads to fragment ions that are driven by the Coulomb repulsion between the charged fragments, releasing potential energy in the form of translational energy of the fragment ions. To the best of our knowledge no report has been published of fast intramolecular reactions in such intense fields. 

We report here results of intense field experiments that we have conducted to probe intramolecular reactions involving proton migration in a series of linear alcohols, methanol, ethanol, propanol, hexanol, and dodecanol. The peak laser intensities employed in our studies were as high as 8$\times$10$^{15}$ W cm$^{-2}$.  Quite unexpectedly, we find that proton migration leading to bond formation occurs within the multiply charged parent molecular ion that is formed by optical field induced tunnel ionization. Moreover, such proton migration occurs much before the Coulomb repulsion makes the fragment ions fly apart. The fast rearrangement leads to formation of unusual products of dissociative ionization of alcohols, namely H$_2^+$ ions that are formed with substantial kinetic energy in the center-of-frame (cm) of the molecule. This ultrafast re-arrangement reaction is seen to be a feature that is common to the fragmentation dynamics of all the linear chain alcohols under similar fields. Results of experiments that we conducted by changing the laser polarization direction indicate that the rearrangement occurs within the 100 fs pulse duration that we used. The short intense light field coaxes molecular rearrangement in addition to causing molecules to undergo fragmentation into ions. Analysis of the fragmentation pathways of the multiple charged alcohols has been carried out using {\it ab initio} quantum computational techniques, and results indicate that H$_2^+$ formation is a preferred pathway. Unlike rearragements that occur in weak-fields, the intense optical field in our experiments coaxes the system to follow the minimum-energy pathway such that hydrogen molecule ions are formed on a time scale that is shorter than that required for the fragment ions from the Coulomb explosion of the multiply charged alcohols to fly apart. 

Our experiments were carried out using a linear, two-field time-of-flight spectrometer (TOF) that has been described elsewhere \cite{expt}. Briefly, we used a Ti:sapphire laser system employing chirped pulse amplification that produces 806 nm wavelength pulses of energy up to 55 mJ/pulse, of 100 fs duration at 10 Hz repetition rate. In the measurements we report here, pulses with energies upto 4 mJ/pulse were focused using a biconvex lens, of 15 cm focal length, in an ultra-high vacuum chamber that is pumped down to a base pressure of 2$\times$10$^{-10}$ Torr. Typical peak intensities in the focal spot were in the range 10$^{15}$-10$^{16}$ W cm$^{-2}$. Use of a 3 mm aperture in front of the TOF spectrometer ensured that our measurements were conducted in the intensity selective mode. The Rayleigh range in these measurements was 6 mm. Linear chain alcohol molecules of interest were effused into the chamber (after degassing by means of several freeze-pump-thaw cycles in a clean, greaseless vacuum line) such that typical operating pressures were in the range of 1-8$\times$10$^{-8}$ Torr. Ions formed in the  interaction region were electrostatically extracted, with nearly unit efficiency, into our TOF spectrometer. A fast photodiode signal was used for the start pulses of our TOF spectrum, while the stop signals were from  the channeltron ion detector used in the pulse counting mode. TOF spectra were measured with a 1 GHz-bandwidth digital oscilloscope along with the waveform for each laser pulse, and data were recorded on the computer using a fast programmable  bus. We also measured the laser pulse intensity along with the arrival time information, such that pulse-to-pulse intensity variations were accounted for, and data could be appropriately selected so as to eliminate intensity fluctuations larger than $\pm$5\%.

Figure 1 a) shows part of a typical TOF spectrum obtained when ethanol molecules were irradiated with pulses of peak intensity 8$\times$10$^{15}$ W cm$^{-2}$. We have recently presented a study of the polarization dependence of the field-induced fragmentation dynamics of linear chain alcohols \cite{jcp-alcohol}. Here we focus only on the proton migration reaction. Along with the multiply charged  molecular ion peaks the TOF spectrum offers evidence for multiply charged atomic fragments, like C$^{2+}$. In such cases, the large kinetic energy release (KER) in the fragmentation event manifests itself in TOF spectra in the form of clear splitting in the arrival times of ions as the fragments that are initially scattered in a direction towards the detector reach early, and are labeled as forward scattered ions (marked $f$ in the figure), while those that are initially formed in the opposite direction reach the detector later, and are denoted backward scattered ions (marked $b$ in the figure). The time difference between the forward and backward ions provides a measure of the kinetic energy release that accompanies a particular ion formation channel. 

We draw attention to the not-insubstantial signal that is observed in Fig. 1 a) at flight times of 0.75 $\mu$s, corresponding to H$_2^+$ molecular ions; Fig. 1 b) depicts this somewhat more clearly. The H$_2^+$ peak shows clear features on either side of the central peak due to the forward and backward scattered ions, which indicates that the ions corresponding to this feature are formed with substantial kinetic energy release in the cm frame. This is an unexpected feature since in TOF spectra one does not expect to find {\em molecular} ions with large energy in the cm frame. Fig. 1 a) also shows a typical spectrum of ion signals obtained at base pressure, when the alcohols are not introduced. At the lower end of the 10$^{-9}$ Torr range, trace H$^+$ ions that are desorbed from stainless steel are observed but we note that there is no evidence for H$_2^+$ ions in the background after our vacuum chamber has been thoroughly degassed by baking for prolonged periods of time. 

H$_2^+$ ions are only observed once alcohols are introduced in the chamber. An important question that obviously needs to be addressed is whether these H$_2^+$ ions result from an intermolecular reaction, where H$^+$ formed from one alcohol molecule interacts with an hydrogen-atom fragment from another molecule within the laser focal volume, or is their formation due to intramolecular reactions? To decipher this we measured the H$_2^+$ signal as a function of alcohol pressure. At pressures below 6$\times$10$^{-8}$ Torr the ion signal that was obtained showed, very clearly, a linear dependence.  The H$_2^+$ yield deviated from linear dependence at higher pressures, indicating the onset of bimolecular collisions. In all our measurements care was taken to use very large extraction fields so as to ensure that ion collection efficiency was close to unity, even for energetic ions.

Although the measurements that we have described so far indicate that H$_2^+$ ions are formed in our experiments by a unimolecular process following an intramolecular rearrangement, unambiguous confirmation of this was obtained from polarization dependent studies that we conducted. Figure 1 b) shows the spectrum that we measured when the plane polarized laser light was perpendicular to the TOF axis as compared to when the plane of polarization was made parallel to the TOF axis. As can be seen, the forward and back-scattered peaks are suppressed with the use of perpendicularly polarized light. Firstly, this enables us to deduce that the energetic H$_2^+$ ions are formed by an intramolecular rearrangement. If they were formed from a bimolecular reaction involving the interaction of a proton with another alcohol molecule, the process would certainly not be expected to show a dependence on the polarization of the ionizing laser pulse. Moreover, it would exhibit a quadratic dependence on gas pressure, contrary to observations. Secondly, the data shown in Fig. 1 b) indicate that the rearrangement reaction occurs within the duration of a single laser pulse, and is driven by the strong optical field. 

We also note that if the formation of H$_2^+$ were a consequence of molecular reorientation, possibly in an excited ionic state that evolved from the Franck-Condon region of the neutral ground state, the H$_2^+$ yield would, again, not be expected to show a dependence on laser polarization. 

As mentioned above, H$_2^+$ ions that are formed also possess substantial kinetic energy in the center-of-mass frame. Figure 2 shows the kinetic energy spectrum that is deduced from the forward and backward scattered peaks in TOF spectra that we measured for methanol precursors.  The TOF spectra were used to obtain the energy released in the cm frame using the following equation derived from simple kinematics \cite{tof-ker}
\begin{equation}
\delta T = {{2(2mU_o)^{1/2}}\over{qE_s}},
\end{equation}
where $\delta T$ is the time separation between the forward and backward scattered peaks, $U_o$ is the energy release in the center-of-mass frame, $m$ is the mass of the ion, $q$ is the ionic charge, and $E_s$ is the ion extraction field used in our TOF spectrometer. There is a low-energy component that essentially arises from those fraction of molecules that are not perfectly aligned along the laser polarization vector and do not undergo fast rearrangements. In such cases H$_2^+$ can be formed from the plethora of avoided crossings between field-dressed potential energy surfaces on timescales that are not necessarily less than 100 fs. Most unexpectedly, energetic hydrogen molecule ions are also produced, with energies that extend up to $\sim$12 eV. The H$_2^+$ kinetic energy arises from the Coulomb repulsion from the charge on the rest of the multiply-charged alcohol molecule. Since a most probable value as high as 5 eV is observed in our measurements (and such values are typical of those expected upon Coulomb explosion of multiply charged molecules \cite{physrep}), the clear implication is that the molecular reorientation occurs well before the Coulomb explosion happens. 

In order to gain further insight, we have made minimum energy path calculations on multiply charged methanol using an {\it ab initio} quantum chemical  method \cite{hyper} to determine the fragmentation path for the linear chain alcohols under consideration here. Self-consistent field Hartree-Fock computations were carried out using a 3-21G basis. Though these calculations are not rigorous enough for deriving the detailed energetics of the fragmentation reaction, they do provide adequately-reliable qualitative insight into the main features of the reaction pathway. Using the steepest-descent method we find that the multiply charged (doubly and triply charged) methanol molecule reorients to form the H$_2$ moiety that separates from the rest of the charged molecule. In Fig. 3 we show snap-shot pictures of the minimum energy path that illustrate this.  This feature was found to be apparent in all the linear chain alcohols used in our experiments. Note how the two hydrogen atoms that are denoted H$_A$ and H$_B$ take part in the rearrangement so as to give rise to formation of a H$_A$-H$_B$ bond in panel f. The novel feature of this reorientation reaction is that it involves an alcohol ion that is multiply charged by the action of a laser field whose intensity is large enough to lead one to intuitively postulate that Coulomb explosion is the only possible pathway that is open. The rearrangement reaction that is depicted in Fig. 3 is coaxed by the strong optical field. The sequence of events that occur are the following: i) at the rising edge of the laser pulse, the C-H axis of the alcohol molecule becomes aligned with respect to the laser light's electric field vector, ii) the aligned molecule undergoes multiple ionization, and iii) also undergoes reorientation so as to form the H$_2^+$ moiety that Coulomb explodes from the remaining part of the charged molecule. 

In summary, we have observed ultrafast intramolecular rearrangement leading to bond formation upon intense field irradiation of a series of linear alcohol molecules in the gas phase. The rearrangement process gives rise to an unusual ionic fragment, H$_2^+$, that is  formed with substantial amount of kinetic energy (a most probable value of about 5 eV in the center-of-mass frame). We have established that H$_2^+$ is formed by a unimolecular reaction. The results of polarization dependence measurements that we have made indicate that the rearrangement occurs within the duration of the laser pulse and is coaxed by the strong light field.

We gratefully acknowledge partial financial support for our Terawatt laser from the Department of Science and Technology.

\begin{figure}
\caption{a) Typical TOF spectra observed when a linear chain alcohol, ethanol in this case, is subjected to linearly-polarized optical fields of intensity 8$\times$10$^{15}$ W cm$^{-2}$. The polarization vector kept was parallel to the spectrometer axis. Similar spectra were also observed for methanol, propanol, hexanol, and dodecanol. Only the fastest arrival times are shown, where ions with low mass to charge ratio ions appear, in order to focus attention only on the H$_2^+$ ion. $f$ and $b$ denote forward and backward scattered ions (see text). The spectrum obtained with same laser intensity and number of laser shots at background pressure ($\leq$1$\times$10$^{-9}$ Torr) is also shown. b) Spectra observed with the plane of polarization parallel and perpendicular to the spectrometer axis. Note the disappearance of forward and backward components in the latter case. The extraction voltages used to measure spectra shown in a) and b) were slightly different, hence the small difference in the flight time corresponding to H$_2^+$ ions.}
\end{figure}

\begin{figure}
\caption{Distribution of kinetic energies released (KER) upon formation of H$_2^+$ ions. The energy values are in the center-of-mass frame.}
\end{figure}

\begin{figure}
\caption{Snapshots of the reaction pathway followed by a doubly-charged methanol ion, CH$_3$OH$^{2+}$ computed by the steepest descent method using an {\it ab initio} Hartree-Fock, self-consistent-field method (see text). Note the rearrangement that leads to formation of a molecular bond between atoms denoted H$_A$ and H$_B$.}
\end {figure}

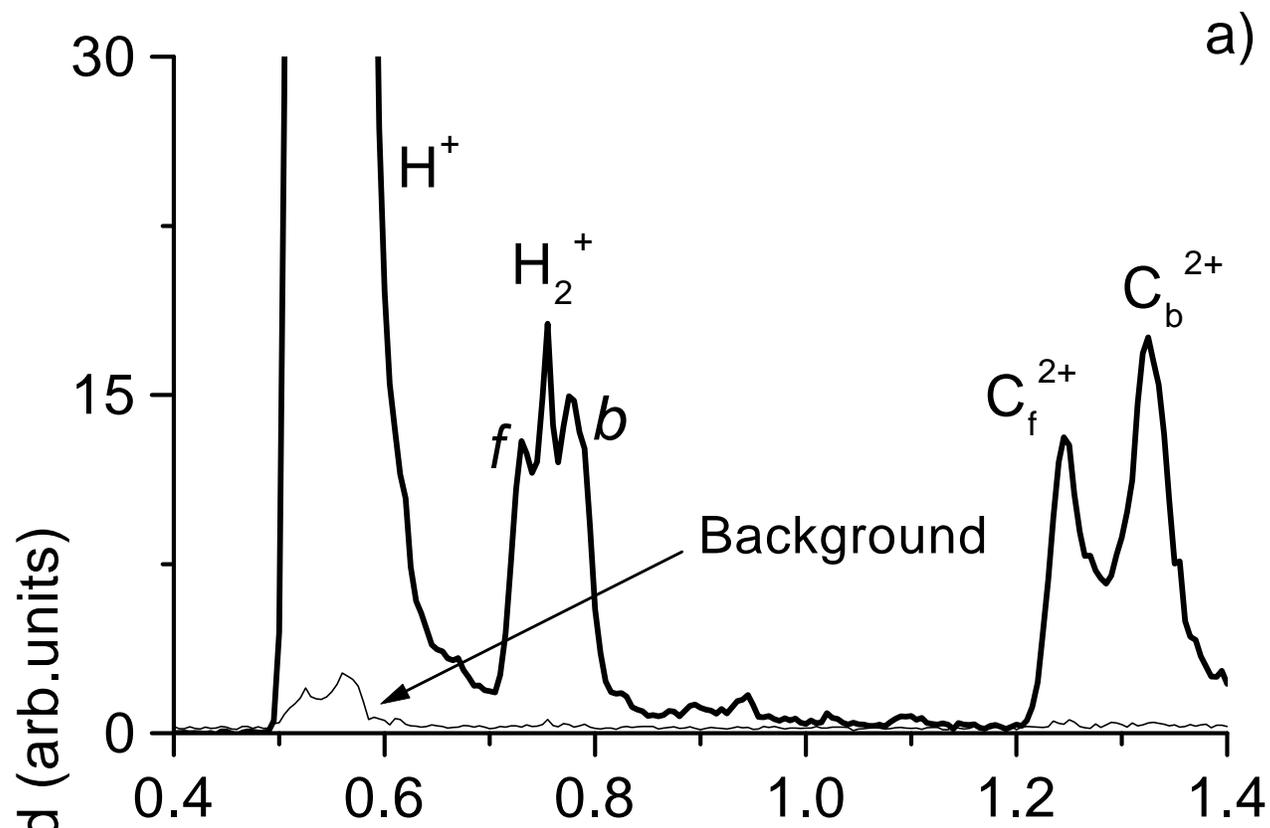
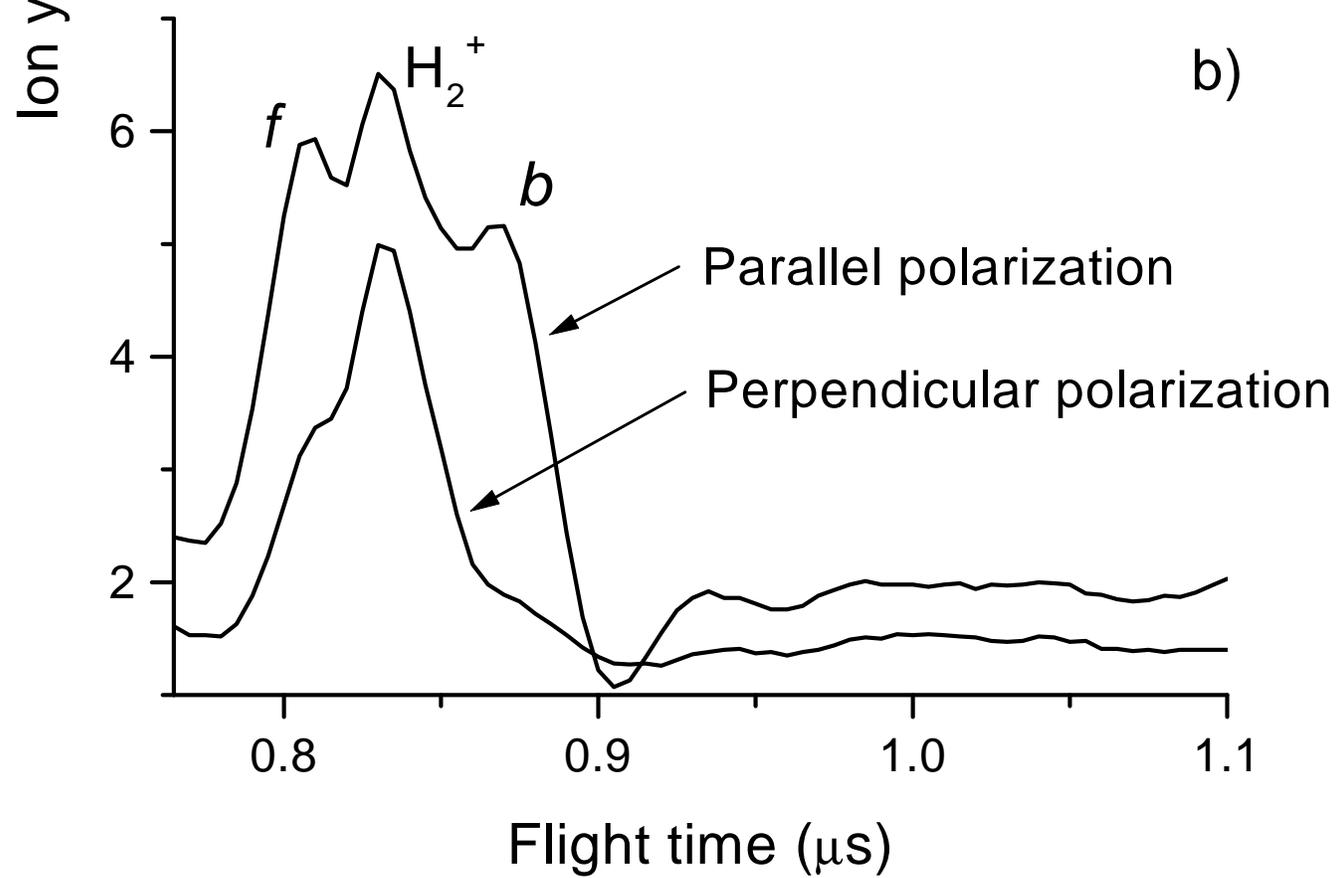
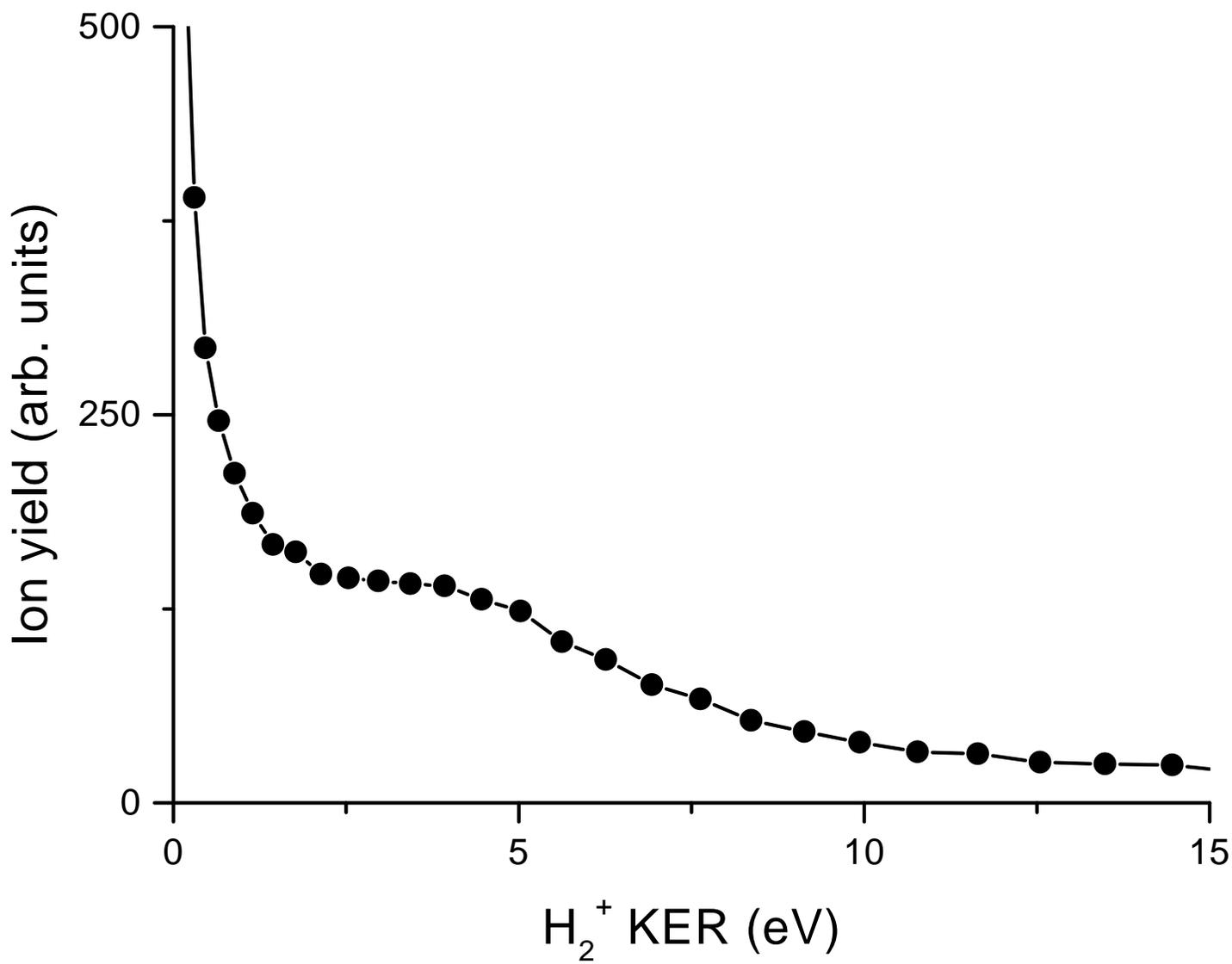
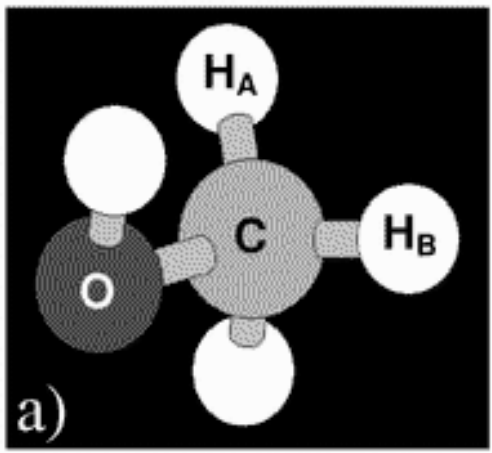
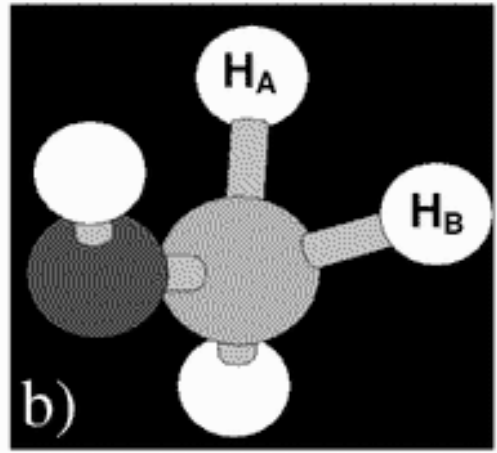
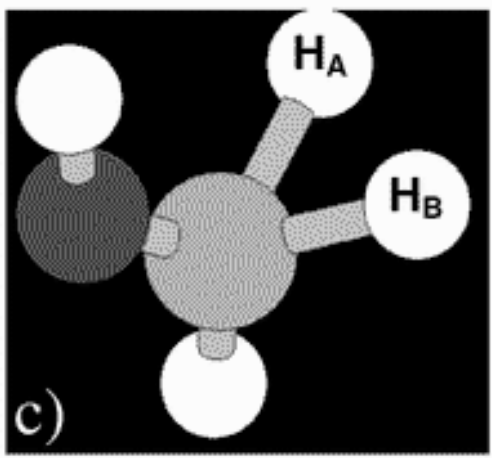
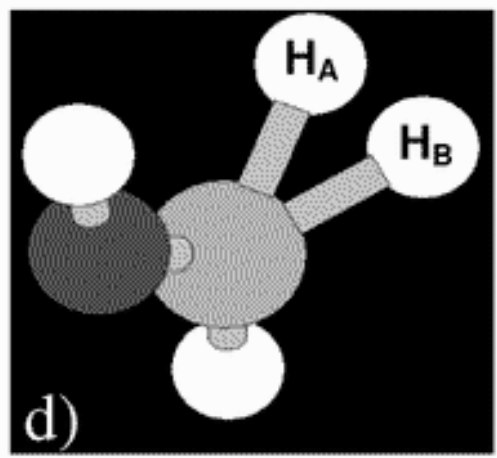
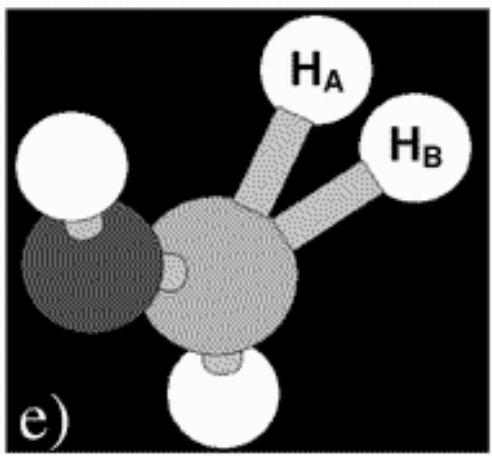
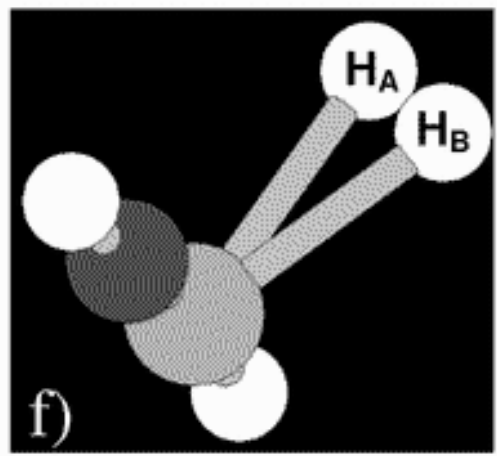
\end{document}